\documentclass[showpacs,amsmath,
reprint,
aip,apl]{revtex4-1}

\usepackage{graphicx}
\usepackage{amsmath}

\begin{document}

\title
{Enhancement of carbon nanotube photoluminescence by photonic crystal nanocavities}
\author{R. Watahiki}
\author{T. Shimada}
\affiliation{Institute of Engineering Innovation, 
The University of Tokyo, Tokyo 113-8656, Japan}
\author{P. Zhao}
\author{S. Chiashi}
\affiliation{Department of Mechanical Engineering, 
The University of Tokyo, Tokyo 113-8656, Japan}
\author{S. Iwamoto}
\author{Y. Arakawa}
\affiliation{Institute of Industrial Science, 
The University of Tokyo, Tokyo 153-8505, Japan}
\author{S. Maruyama}
\affiliation{Department of Mechanical Engineering, 
The University of Tokyo, Tokyo 113-8656, Japan}
\author{Y. K. Kato}
\email[Corresponding author: ]{ykato@sogo.t.u-tokyo.ac.jp}
\affiliation{Institute of Engineering Innovation, 
The University of Tokyo, Tokyo 113-8656, Japan}

\begin{abstract}
Photonic crystal nanocavities are used to enhance photoluminescence from single-walled carbon nanotubes. Micelle-encapsulated nanotubes are deposited on nanocavities within Si photonic crystal slabs and confocal microscopy is used to characterize the devices.  Photoluminescence spectra and images reveal nanotube emission coupled to nanocavity modes. The cavity modes can be tuned throughout the emission wavelengths of carbon nanotubes, demonstrating the ability to enhance photoluminescence from a variety of chiralities.
\end{abstract}

\maketitle

Single-walled carbon nanotubes (SWCNTs) are unique nanoscale infrared light emitters, and  various potential applications have been suggested in broad areas including physics, optoelectronics, and biology. It has been shown that micelle-encapsulated nanotubes can be single photon sources at low temperatures.\cite{Hogele:2008} By integrating individual SWCNTs into field-effect transistors, light emission can be induced electrically.\cite{Misewich:2003, Chen:2005, Mann:2007} The infrared wavelengths and biocompatibility makes them ideal for fluorescence imaging in biological systems.\cite{Cherukuri:2004, Leeuw:2007, Welsher:2009} Unfortunately, their quantum efficiencies are typically fairly low,\cite{O'Connell:2002, Chen:2005, Crochet:2007, Miyauchi:2009} putting limitations on their use. 

A possible pathway for enhancing light emission from nanotubes is to place them in an optical cavity. There have been reports on emission enhancement by planar cavities formed by metallic mirrors\cite{Xia:2008} and dielectric mirrors,\cite{Gaufres:2010oe} as well as half cavities with a single metallic mirror and a solid immersion lens.\cite{Walden-Newman:2012} Compared to these conventional planar cavities, photonic crystal (PC) cavities offer a significant advantage for enhancing photoluminescence (PL) of nanomaterials. Their small mode volumes and high quality factors result in a strong Purcell effect, and they also allow redirection of the emission for more efficient detection.\cite{Noda:2007, Fujita:2008} For example, they have been used to enhance light emission from self-assembled quantum dots,\cite{Yoshie:2001, Noda:2007} colloidal quantum dots,\cite{Fushman:2005, Wu:2007} and nitrogen-vacancy centers in diamond.\cite{Englund:2010, Wolters:2010}

Here we report on enhancement of PL from SWCNTs by Si PC nanocavities. Fluorescent nanotube solutions are directly deposited on the devices, and they are characterized with a home-built laser-scanning confocal microscope. We observe sharp peaks at wavelengths much longer than Si PL that are localized at the cavity location, and attribute them to nanotube PL coupled to PC cavities. We have observed quality factors as high as 3800 and PL enhancement factors larger than $50$. We also demonstrate tuning of the cavity resonance throughout the nanotube emission wavelengths by changing the PC lattice constant $a$.

\begin{figure}
\includegraphics{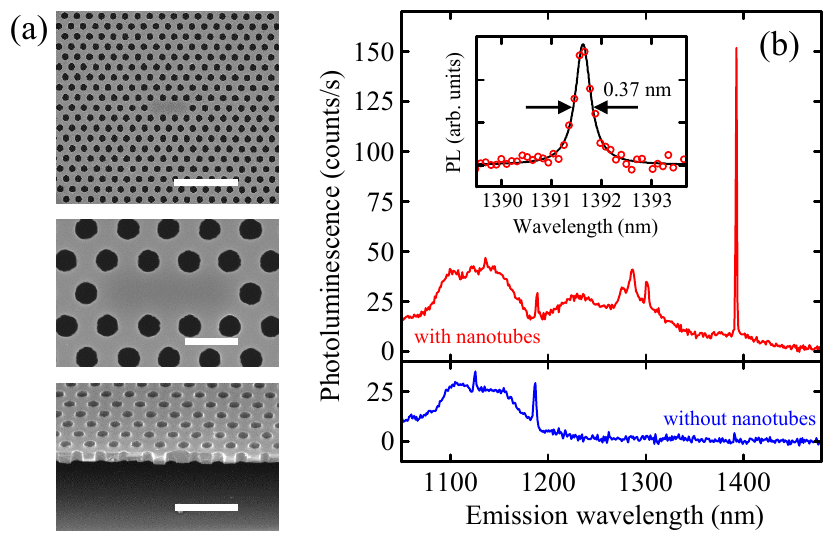}
\caption{\label{fig1}
(a) Scanning electron micrographs of as-fabricated devices with $a=380$~nm. The top panel shows a plan view, and the middle panel is an enlarged view of the cavity. The bottom panel shows a cross section of the slab structure. The scale bars are 2.0, 0.5, and 1.0~$\mu$m for the top, middle, and bottom panels, respectively. 
(b) Top and bottom panels show PL spectra taken at the cavity with and without nanotubes, respectively, for devices with $a=380$~nm and $r=100$~nm. Inset shows a higher resolution spectrum of the fundamental mode. Circles are data and the solid line is a Lorentzian fit. An excitation laser wavelength of 750~nm and a power of 0.5~mW are used.
}\end{figure}

The PC nanocavities are fabricated from silicon-on-insulator wafers with a 200-nm-thick top Si layer and a 1000-nm-thick buried oxide layer. Electron beam lithography is performed to define the hexagonal lattice PC pattern, and a Bosch process in an inductively-coupled plasma etcher forms the air holes in the top Si layer. We use the ``L3'' cavity, which consists of three missing holes in a line. The holes are actually designed as regular hexagons with side length $r$ to reduce the data size used by the electron beam writer, but the lithography and etching processes results in circular holes.\footnote{The actual radius of the holes as measured by a scanning electron microscope is larger than $r$ by 5 to 10~nm.} In order to form a free-standing slab structure, the buried oxide layer is etched in 20\% hydrofluoric acid for 10 minutes. Scanning electron microscope images of typical devices are shown in Fig.~\ref{fig1}(a).

Micelle-encapsulated nanotube solutions\cite{O'Connell:2002}  are prepared from commercially available SWCNTs produced by the high-pressure carbon monoxide process. The nanotubes are dispersed in water containing 10 wt\% of sodium dodecylbenzenesulfonate, and then the solution is bath sonicated for 10 minutes, tip sonicated for 30 minutes, and ultracentrifuged for 30 minutes at 327,000$g$. We extract the supernatant of the solution for our experiment.

The devices are characterized with a home-built laser-scanning confocal micro-PL system optimized for nanotube fluorescence detection.\cite{Moritsubo:2010, Yasukochi:2011}  An output of a wavelength-tunable continuous-wave Ti:sapphire laser is focused onto the sample by an infrared objective lens with a numerical aperture of 0.8 and a working distance of 3.4~mm. PL is collected through the same lens, and the laser scatter is rejected with a dichroic beam splitter and a long-pass filter. A pinhole corresponding to a 2.7~$\mu$m aperture at the sample imaging plane is placed at the entrance of a 300-mm spectrometer for confocal detection. PL is dispersed with either a 150 lines/mm grating blazed at 1.25~$\mu$m or a 900 lines/mm grating blazed at 1.3~$\mu$m, and detection is performed by a liquid-nitrogen-cooled 512-pixel linear InGaAs photodiode array with pixel widths of 50~$\mu$m.  A coil-driven fast steering mirror allows scanning of the laser spot for collecting PL images. All measurements are done in air at room temperature.

We deposit nanotubes on the devices by putting a drop of nanotube solution on the chip and letting it dry in air on a hot plate at $120^\circ$C. Figure \ref{fig1}(b) shows PL spectra taken at the cavity position for a device with nanotubes and another device without nanotubes. The device without nanotubes show Si PL from around 1100 to 1200~nm with a sharp peak near 1190~nm which we assign  to the 5th mode of the L3 cavity.\cite{Fujita:2008, Iwamoto:2007} In comparison, for the device with nanotubes, we observe broad emission for wavelengths longer than 1200~nm with a very sharp peak near 1400~nm and a few more peaks around 1300~nm. 
The broad emission is consistent with PL from SWCNTs,\cite{O'Connell:2002} while the positions of the additional peaks agree with the other modes of the L3 cavity, with the sharpest peak near 1400~nm being the fundamental mode.\cite{Fujita:2008} Our finite-difference time-domain calculation for this device gives the fundamental mode at 1443~nm, confirming this assignment. As these wavelengths are much longer than Si PL, we attribute these peaks to nanotube emission coupled to the PC cavity modes.

The inset of Fig.~\ref{fig1}(b) shows a higher resolution spectrum of the fundamental mode taken with the 900 lines/mm grating. A Lorentzian fit gives a full-width-at-half-maximum of 0.37~nm corresponding to a quality factor $Q=3800$. Considering that our numerical calculation gives $Q=6100$ for this device, the deposition of nanotubes on PC cavities does not seem to degrade the cavity quality too much. 

\begin{figure}
\includegraphics{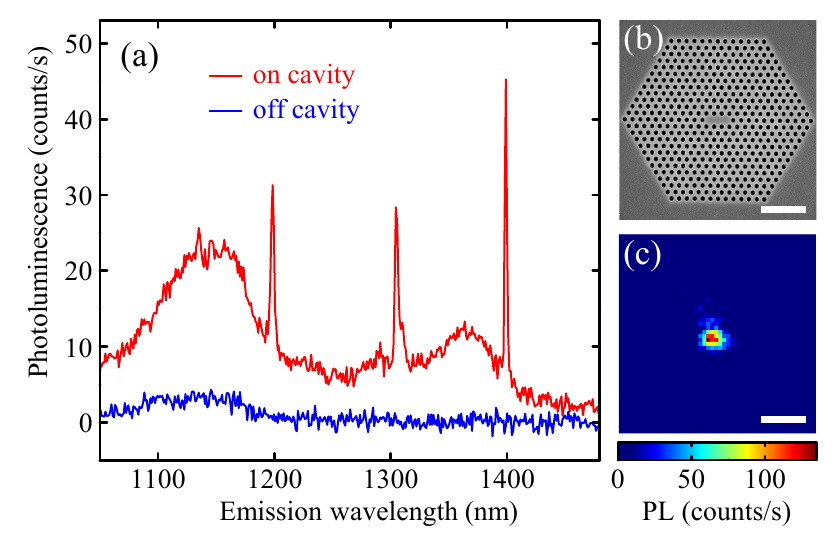}
\caption{\label{fig2}
(a) PL spectra taken on the cavity (red) and off the cavity (blue) for a device with $a=382$~nm and $r=104$~nm.  
Nanotubes have been deposited on this device by spin coating at 1400 rotations per minute. The excitation wavelength is 846~nm and the power is 0.3~mW.
(b) A scanning electron micrograph showing the area of the PL image in (c).   A device with the same $a$ and $r$ as in (a) but without nanotube deposition is used.
(c) A PL image of the same device as shown in (a), taken with a detection window centered at 1395~nm with a width of 5.2~nm. In order to construct this image, the PL counts have been integrated over the 5 pixels corresponding to this spectral width. The scale bars in (b) and (c) are 2~$\mu$m.
}\end{figure}

Coupling between nanotubes and PC cavities can be achieved for much smaller quantities of nanotubes. For a different device, we use spin coating which results in much less nanotube deposition. With a laser position off the cavity, we barely see any signal within nanotube emission wavelengths [Fig.~\ref{fig2}(a), blue curve]. Even at such a low nanotube density, we can still observe cavity modes when the laser spot is on the cavity [Fig.~\ref{fig2}(a), red curve]. 

The PL enhancement factor can be obtained by taking the ratio of emission intensities for these two spectra. Averaging the off-cavity spectrum in a 40-nm window centered around the fundamental mode results in 0.2~counts/s for the background nanotube emission. The enhancement factor calculated using this value would result in a factor of more than 200, but since the background counts are close to zero, we take the noise level as the background for a more conservative estimate. The root-mean-squared value of the same spectral window is 0.8~counts/s, resulting in a lower bound for the enhancement factor of about 50.  

On this device, we explore the spatial extent of the nanotube emission coupled to the cavity mode. 
For the area shown in Fig.~\ref{fig2}(b), a PL image is taken at an emission wavelength of the fundamental mode for this device [Fig.~\ref{fig2}(c)]. As expected, the enhanced emission is localized at the cavity.  

We note that the nanotubes in our devices are on the surface of the cavity, and the calculated value of squared electric field at the surface is about 40\% of that at the center of the slab thickness for the fundamental mode. It is promising that we observe PL enhancement despite the reduced electric fields, and it may be possible to increase the enhancement with a cavity design where the electric field is maximized at an air hole.\cite{Vuckovic:2001} We also note that the nanotubes are randomly oriented in our devices, resulting in less-than-optimal coupling. Since nanotube emission is polarized along the tube axis,\cite{Hartschuh:2003, Misewich:2003, Lefebvre:2004, Moritsubo:2010} we should be able to obtain stronger effects if nanotubes are aligned to the cavity polarization. 

\begin{figure}
\includegraphics{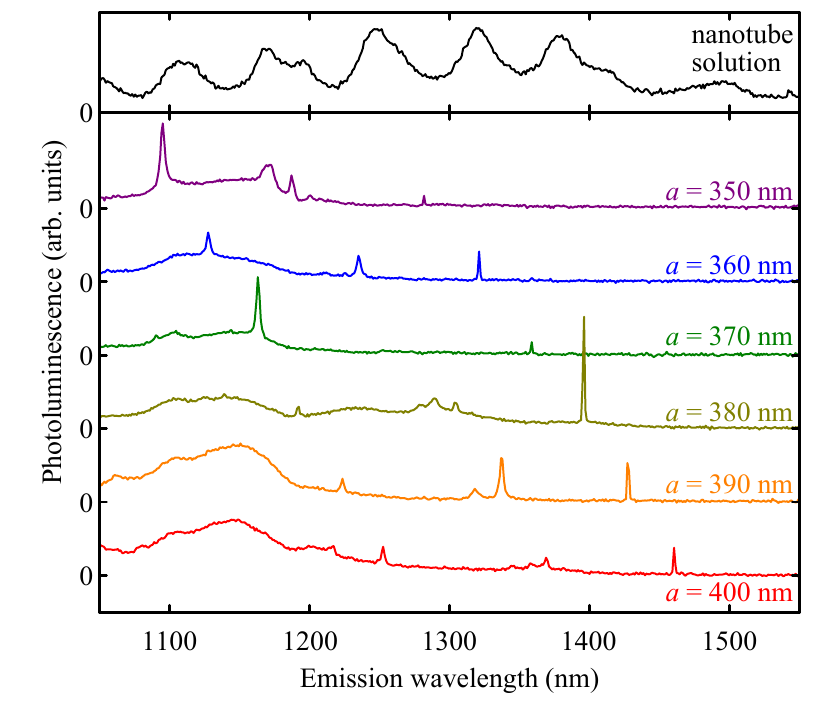}
\caption{\label{fig3}
The top panel shows a PL spectrum from nanotube solution taken with an excitation wavelength of 774~nm. The bottom panel shows PL spectra for devices with $r=100$~nm and $a$ ranging from 350~nm to 400~nm, taken with an excitation wavelength of 750~nm and a power of 0.5~mW. The curves are offset for clarity.
}\end{figure}

Finally, we demonstrate wavelength tuning of the cavity modes throughout the emission spectrum of nanotubes. In the top panel of Fig.~\ref{fig3}, a PL spectrum of the nanotube solution is shown. There are a number of peaks corresponding to different chiralities,\cite{O'Connell:2002, Bachilo:2002} spanning a broad spectral range. We have fabricated devices with $a=350$~nm to $a=400$~nm, and deposited nanotubes in the same manner as in the device shown in Fig.~\ref{fig1}(b). In the bottom panel of Fig.~\ref{fig3}, we show PL spectra from these devices. The cavity modes redshift with increasing $a$, consistent with previous work on Si PL enhancement.\cite{Fujita:2008, Iwamoto:2007} Such redshifts combined with different modes of the cavity allow tuning throughout the broad spectrum of the nanotube PL.

In summary, we have demonstrated enhancement of PL from SWCNTs using Si PC nanocavities, by a factor of at least 50. For the fundamental mode of the L3 cavity, $Q$ can be as high as 3800 with nanotubes deposited on the devices, showing that SWCNT emission can be efficiently coupled to PC nanocavity modes without significantly degrading their quality. It is possible to tune the cavity modes throughout the emission spectrum of nanotubes by changing the lattice constant of the PC. Matching a mode to one of the emission peaks of nanotubes should allow selective enhancement of PL for a particular chirality. By combining ultra-high-$Q$ PC nanocavities \cite{Song:2005} with nanotubes that show optical gain, \cite{Gaufres:2010apl} it may be possible to achieve lasing of carbon nanotubes.

\begin{acknowledgments}
We thank S. Nakayama, R. Ohta, T. Tanabe, and Y. Takahashi for helpful discussions. Work supported by SCOPE, KAKENHI (21684016, 22226006, 23104704, 24340066, 24654084), SCAT, Asahi Glass Foundation, Casio Science Promotion Foundation, and KDDI Foundation, as well as the Global COE Program, Photon Frontier Network Program, and Project for Developing Innovation Systems of MEXT, Japan. The devices were fabricated at the Center for Nano Lithography \& Analysis at The University of Tokyo.
\end{acknowledgments}

\bibliography{CNTandPC}

\end{document}